\documentclass[12pt]{article}
\usepackage{amsmath,latexsym,epsfig,amssymb}

\def\l{\left}
\def\r{\right}
\def\nn{\nonumber}

\def\ord#1{{\mathcal O}\l(#1\r)}

\def\bea{\begin{eqnarray}}
\def\eea{\end{eqnarray}}

\def\Tr{\mathrm{Tr}}
\def\Re{\mathrm{Re}}

\def\gev{\mbox{ GeV}}
\def\mev{\mbox{ MeV}}

\newcommand\be{\begin{equation}}
\newcommand\ee{\end{equation}}

\def\ma[#1,#2,#3,#4]  {{\left( \matrix{ #1  & #2 \cr
                                        #3  & #4 \cr } \right)}}

\begin{document}
\title{
{\vspace{-1cm} \normalsize
\hfill \parbox{40mm}{CERN/TH-99-197}\\
\hfill \parbox{40mm}{CPT-99/PE.3856}\\
\hfill \parbox{40mm}{FTUV/99-49 \\IFIC/99-51}}\\[25mm]
Finite-size scaling of the quark condensate in quenched lattice QCD}
\author{
Pilar Hern\'andez\footnote{On leave from Departamento de F\'{\i}sica Te\'orica, Universidad de Valencia.},
Karl Jansen\footnote{Heisenberg Foundation Fellow}
$\;$ and Laurent Lellouch\footnote{On leave from Centre de Physique 
Th\'eorique, CNRS Luminy, F-13288 Marseille Cedex 9, France.}\\
CERN, 1211 Geneva 23, Switzerland}
%
\maketitle

\begin{abstract}

  We confront the finite volume and small quark mass behaviour of the scalar
  condensate, determined numerically in quenched lattice QCD using
  Neuberger fermions, with predictions of quenched chiral perturbation
  theory.  We find that quenched chiral perturbation theory
  describes the numerical data well, allowing us to extract the
  infinite volume, chiral limit scalar condensate, up to a
  multiplicative renormalization constant.
\end{abstract}

\pagebreak

{\em Introduction}\\

Chiral symmetry breaking plays a central role in our comprehension of
low energy QCD and understanding it from first principle calculations is
of great importance. One of the cleanest ways of determining a
condensate associated with the breaking of a global symmetry is
through a finite-size scaling analysis. This technique has proved very
successful in the study of scalar ${\rm O(N)}$ models~\cite{phi4}. For chiral
symmetry breaking in QCD, this would correspond to placing the system
in a box and studying the scaling of the scalar condensate as a
function of the volume $V$ and of the quark mass $m$ as the limit of
restoration of chiral symmetry is approached ($m\to 0$, $V$ finite).

The very small quark mass limit of QCD is expected to be well
described by the lowest orders of chiral perturbation theory
($\chi$PT), which predict how the restoration of chiral symmetry takes
place in a finite volume, as a function of the quark mass \cite{ls}.
The only free parameter entering the
leading order contribution in the chiral expansion is the infinite volume
quark condensate.  Thus, a comparison of the mass and volume
dependence of the finite volume quark condensate with the predictions
of $\chi$PT provides a very powerful test of the hypothesis of
spontaneous chiral symmetry breaking and permits an extraction of the
infinite volume scalar condensate $-\Sigma$.
Such a study requires, however, a good control over the chiral properties of the
theory, which is difficult to achieve with traditional formulations of
fermions on the lattice.

The situation is different, however, when Dirac operators that
satisfy the Ginsparg--Wilson (GW)
relation ~\cite{gw,hasen2} are considered. Actions constructed from such
operators have been shown to have an exact lattice chiral
symmetry~\cite{luscher}. This symmetry ensures that the relations
implied by chiral symmetry in the continuum, hold also on the lattice
at finite lattice spacing $a$~\cite{neu}--\cite{chandra}.  For a
review of the GW relation and its implications, we refer to
\cite{nieder}.

A particular realization of an operator satisfying the GW relation,
which we will be using here, has been proposed by
Neuberger \cite{neu}:
\begin{equation}
\label{defneuop}
 D_{\rm N} \equiv [m + (1+s) (1 - \gamma_5 Q (Q^2)^{-1/2})],
\end{equation}
where $Q \equiv c_0\gamma_5 (1 + s - D_{\rm W})$,  $D_{\rm W}$ is the
Wilson Dirac operator, and the factor $c_0$ is a convenient
normalization to keep the spectrum of $Q^2$ bounded by 1. 
The parameter $s$ satisfies $|s|<1$ and $m$ is the bare quark mass.

$D_{\rm N}$ satisfies the GW relation at zero quark mass.  In contrast
to the standard Wilson formulation, the breaking of the 
chiral symmetry is soft, i.e. only due to the quark mass term.  This
opens the possibility to confront finite volume simulations with
finite-size scaling predictions in the regime of restoration of the 
chiral symmetry.

The complexity of the operator $D_{\rm N}$ renders its numerical
treatment very demanding. We therefore restrict to the quenched
approximation. The predictions of $\chi$PT must then be modified
 to take into account the effect of quenching.  The finite
size scaling of the quark condensate has recently been worked out
using the framework of quenched chiral perturbation theory (q$\chi$PT) \cite{osborn}. In particular, analytical expressions for this
scaling have been obtained in sectors of fixed topology. Operators satisfying the GW relation
also satisfy an index theorem \cite{luscher}.  Thus, by computing the
eigenvalues of $D_{\rm N}$ at zero quark mass and identifying the zero
modes, a clean separation of different topological sectors can be
achieved, which is not possible with other formulations of lattice 
fermions. As we will see
below, using the q$\chi$PT results in fixed topological sectors to interpret our numerical data
proves very useful. A preliminary account of this work was presented at 
Lattice 99.

\vspace{0.8cm} 

{\em Light quarks on a torus}\\
To study the volume dependence of the scalar condensate, we work on a
four-dimensional torus of volume $L^4$. Under the assumption that
chiral symmetry is spontaneously broken, a description of QCD in
terms of a chiral Lagrangian should be a good approximation at
momenta $p \ll 4 \pi F_{\pi}$. To lowest order in $p/4 \pi F_{\pi}$ and in the
quark mass, this Lagrangian is given by
\begin{eqnarray}
{\cal L} = \frac{F_{\pi}^2}{4} \;\Tr[\;\partial_\mu U^\dagger(x) \;
\partial_\mu U(x)] - \Sigma \; \Re\;\Tr[M 
e^{-i \theta/N_f} U]
\label{lagrange1}
\end{eqnarray}
where $U(x) = \exp[ i 2 \Pi(x)/F_\pi] \in SU(N_f)$, $\Pi(x)$ being the
pion fields; $M$ is the quark mass matrix, which we take to be
proportional to the identity matrix (i.e. $M=m I$), and $-\Sigma$ is
the infinite volume and zero quark mass scalar condensate. In
eq.~(\ref{lagrange1}) we have included the expected $\theta$ angle
dependence.

Let us now consider the regime
\begin{eqnarray}
M_{\pi}\ll 1/L \ll F_{\pi}\ ,
\end{eqnarray}
where $M_\pi^2=2m\Sigma/F_\pi^2$ to leading order in $\chi$PT. In this regime, the partition
function is dominated by the zero mode of the $U(x)$ field
\cite{ls}, since the action of the non-zero modes has a
kinetic contribution that goes like $F^2_{\pi} L^2 \gg 1$. The
partition function then reduces, to leading order, to an integral
over the $SU(N_f)$ group manifold:
\begin{eqnarray}
Z = \int_{SU(N_f)}  d U_0 \; e^{ V \Sigma\Re \; \Tr[ M e^{-i \theta/N_f} U_0]}\ ,
\end{eqnarray}
where $U_0$ is the global mode. We can also define the partition
function restricted to fixed topology by Fourier transforming in
$\theta$ \cite{ls}:
\begin{eqnarray}
Z_\nu &=& \int^{2 \pi}_0 \frac{d \theta}{2\pi} 
\int_{SU(N_f)}  d U_0 \; e^{-i \theta \nu}\;
\exp[V \; \Sigma \; \Re\Tr[ M e^{i \theta} U_0]]\nonumber \\
&=&\int_{U(N_f)} d U_0 \; \det(U_0)^\nu \; \exp[V \; \Sigma\; \Re\Tr[ M U_0]].
\end{eqnarray}
These integrals and their derivatives with respect to the quark mass
have been known for a long time. For details see \cite{ls}. 

In our case, however, we are interested in the quenched approximation. 
Recently a similar reasoning has been applied to quenched QCD. The
main difference in the quenched case is that the chiral symmetry
group is no longer $SU(N_f)_L\times SU(N_f)_R\times U(1)$, 
but a graded Lie group $U(1|1)_L\times U(1|1)_R/U_A(1)$.
According to \cite{osborn}, the partition
function for fixed topology is then given by
\begin{eqnarray}
Z_\nu = \int_{U(1|1)} d U_0 \; s\det(U_0)^\nu \; \exp[V \;\Sigma \;\Re
s\Tr[ M U_0]].
\end{eqnarray}
This integral has been computed analytically in terms of Bessel
functions \cite{osborn}. By differentiating its logarithm with respect to the
quark mass, $m$, the quark condensate for fixed topology is found to
be
\begin{eqnarray}
\Sigma_\nu = \Sigma\; z \; [ I_\nu(z) K_\nu(z) + I_{\nu+1}(z)
K_{\nu-1}(z)] + \Sigma\frac{\nu}{z},
\label{qpt}
\end{eqnarray}
where $z \equiv m \Sigma V$ and $I_\nu(z), K_\nu(z)$ are the modified Bessel
functions.

This formula summarizes the scaling of the quark condensate in a
periodic box with the volume and quark mass in the small $m \Sigma V$
limit, as a function of only one non-perturbative parameter: the
infinite volume condensate $-\Sigma$. For fixed volume, the limit as
$m \rightarrow 0$ is given by
\begin{eqnarray}
\Sigma_{\nu=0} & = & m \;\Sigma^2\; V\; 
\l( 1/2 - \gamma + \log{2} - \log{m \Sigma V}+
\ord{m \Sigma V \log{m \Sigma V}}^2\r)\nn\\
\Sigma_{\nu=\pm 1} & = &
 \frac{1}{m V} + \frac{1}{2} m \Sigma^2 V \l(1+\ord{m \Sigma V \log{m \Sigma V}}^2 \r).
\label{exp}
\end{eqnarray}
where $\gamma$ is the Euler constant. These results have two
interesting features that we wish to emphasize. First, there is a
divergence $\sim 1/m$  in sectors with topology. From the point of view
of the underlying theory, this is not surprising since it corresponds to the
contribution of the fermionic zero modes.
Note however that these terms do not contain information
about the infinite volume condensate and vanish in the infinite volume limit, as expected. The second interesting feature
is the appearance of a logarithmic enhancement in $\Sigma_{\nu=0}$, which 
is also peculiar to the quenched approximation. This term contains 
information about the infinite volume condensate.         

In principle, by fitting the dependence of the finite volume
condensate in quark mass and volume to Monte Carlo data, we can extract
the infinite volume condensate. However, the naive bare quark
condensate that is measured on the lattice is UV-divergent. A simple 
dimensional analysis of the possible divergences shows that the bare 
scalar  condensate  has a leading cubic divergence. 
One important advantage of Neuberger's operator is that the coefficient of this leading divergence is 
known analytically. It is $6/(1+s)$, for $SU(3)$. 
The cubic divergence can then be subtrated {\rm exactly}. However, 
after this trivial subtraction, the condensate is still divergent and has the 
form:
\begin{eqnarray}
\Sigma^{sub}_{\nu}(a)  \equiv -\;\langle \bar \Psi \;
\Psi \rangle_{\nu} - (\frac{6}{1+s}) \frac{1}{a^3} = C_2 \frac{m(a)}{a^2} + C_1 \frac{m(a)^2}{a} + \Sigma_{\nu},
\label{eq:sigsub}
\end{eqnarray}
where $m(a)$ is the bare lattice mass.  The
constants $C_i$ are not known a priori and have to be determined,
preferably non-perturbatively. The linear divergence proportional 
to $m(a)^2$ is negligibly small 
for the values of the mass and the cutoff we consider in this work.
  However,
the quadratic divergence is not and turns out to be very important 
numerically. The condensate extracted through a
fit of the lattice data to eqs.~(\ref{eq:sigsub}) and (\ref{qpt}), of
course, still requires a multiplicative renormalization to eliminate
a residual logarithmic UV divergence in $\Sigma_{\nu}$. 

After subtracting the unphysical contribution of fermionic zero modes to 
$\Sigma_\nu$,  
the finite volume condensate $\Sigma^{sub}_{\nu}$ vanishes, 
as expected, in the limit of zero
quark mass \cite{hasen3,florida}. Not surprisingly, the power divergences can be separated,
in principle, from the physical contribution to the 
condensate, by a study of the volume dependence of $\Sigma^{sub}_{\nu}$, while
keeping the quark mass small enough to stay in the region of validity of 
$\chi$PT. 

Even with the cubic divergence already subtracted, separating the
physical condensate from the remaining power divergences may not be
easy, in practice, because the statistical errors in these divergences
can hide the small physical contribution. Thus, $\Sigma^{sub}_{\nu}$
must be computed with very good accuracy.  Clearly, the logarithmic
enhacement of $\Sigma_{\nu=0}$ in eq.~(\ref{exp}) could be very helpful in this respect;
however, as we will see, extracting the condensate from the
logarithmic term at zero topology requires much larger statistics than
available to us at this time. We will concentrate instead on the study of the condensate
in the topological charge one (or minus one) sector.

\vspace{0.8cm}

{\em Numerical results}\\ 
For our numerical simulations we work on hypercubic lattices of
size $L^4$ with periodic boundary conditions for both the
gauge and the fermion fields. We work in the quenched approximation
and use standard methods to obtain decorrelated gauge field
configurations. 

In selecting the value of $\beta =6/g_0^2$, some care has to be taken.
On the one hand, the quadratic divergence $\propto 1/a^2$ should not
hide the physical effect. On the other hand, in choosing too small
values of $\beta$ there is the risk that Neuberger's operator falls
into a different universality class \cite{hjl}. Indeed, by computing
the low-lying eigenvalues of Neuberger's operator at $\beta=5.7$ and
$s=0$, we only found eigenvalues ${\rm O(1)}$ and hence no light
physical modes. A scan of the lowest eigenvalue of $Q^2$ as a function
of $s$ showed that $\lambda_{\rm min}(Q^2)$ {\em decreased} with
increasing $s$, contrary to what is expected (and found) at larger
values of $\beta$.

The situation at $\beta=5.85$ appeared to be different, however.  The
values of $\lambda_{\rm min}(Q^2)$ reach a maximum around $s=0.6$,
where the localization properties should also be
optimal~\cite{hjl}.  In accordance, the eigenvalues of Neuberger's
operator became very small, so that there is little doubt that light, physical
modes are present. Since the lattice spacing at $\beta=5.85$ is
$a^{-1}\approx 1.5 {\rm GeV}$ \cite{a} we estimated that the quadratic
divergence term would not hide the physical signal, at least for
reasonable values of the physical condensate.

A technical challenge is the numerical treatment of the square root
appearing in Neuberger's operator. We have chosen a Chebyshev
approximation for this task, which allows us to reach a well
controlled accuracy. In order to avoid any systematic effects in the
values of physical observables, we demand that
\begin{equation}
\| X - Q^2 P_{n,\epsilon}(Q^2)^2 X \|^2 /(2\|X\|)^2
 < 10^{-16}\; .
\label{accuracy}
\end{equation}
In eq.~(\ref{accuracy}) $X$ denotes a random vector and $P_{n,\epsilon}$
denotes a standard Chebyshev approximation of the function $1/\sqrt{x}$
in the range $\epsilon \le x \le 1$. $P_{n,\epsilon}$ is a matrix-valued polynomial of degree $n$, which is constructed through numerically 
stable recursion relations~\cite{numrec}. We require tantamount accuracies
for all inversions. We note in passing that with the requirement of eq. (\ref{accuracy}) also the GW relation itself is satisfied to a similar
accuracy for zero mass.

In order to decrease the degree of the polynomial employed, we have
computed the 11 lowest eigenvalues of $Q^2$ and their corresponding
eigenvectors and have set $\epsilon$ to be the value of the largest.
The contributions of these lowest lying eigenvectors are then treated
exactly and projected out of the operator $Q^2$. Through this
procedure, near-zero modes of $Q^2$ are taken into account
automatically. All eigenvalue computations performed in our work are
based on minimizing the Ritz functional~\cite{cg}.

As pointed out in \cite{florida}, it is advantageous for the
computation of the eigenvalues of Neuberger's operator, and for its
inversion as well, to stay in a given chiral subspace. This is possible
because $D_N^\dagger D_N$ commutes with $\gamma_5$. 

We computed the scalar condensate at several values of the quark mass
using a multiple mass solver \cite{mulmass} on lattices of size $8^4$,
$10^4$ and $12^4$.  We checked through the calculation of the two lowest
eigenvalues of Neuberger's operator to which topological sector each
gauge field configuration belonged.  We then obtained
$\Sigma^{sub}_{\nu}$ by computing
\begin{equation}
\Sigma^{sub}_\nu=
\frac{1}{V}\langle\Tr'\l\{
\frac{1}{D_N}+\frac{1}{D_N^\dagger}-\frac{a}{1+s}\r\}\rangle_\nu,
\label{trace}
\end{equation}
where the trace was performed in the chiral sector opposite to that
with the zero modes and the gauge average was done in a sector of fixed
topology $\nu$. With this definition, we take into account the contribution 
of all the non-zero eigenvalues of $D_N$ to the condensate \footnote{With this 
definition the real eigenvalues at the cut-off level, $m+2/a$, are doubly
counted. Although it is a completely negligible effect, we took it into account.}. In this way, the term $\sim 1/m$ in eq.~(\ref{exp}) is absent. Three gaussian sources and 
standard inverters were used to compute the trace in eq. (\ref{trace}).
Topological charge zero configurations are very rare at larger volumes. For
this reason we did not compute the condensate in this sector on the
$10^4$ and $12^4$ lattices since the statistics we have gathered are
too small. 

\begin{figure}
\vspace{0.0cm}
\begin{center}
\psfig{file=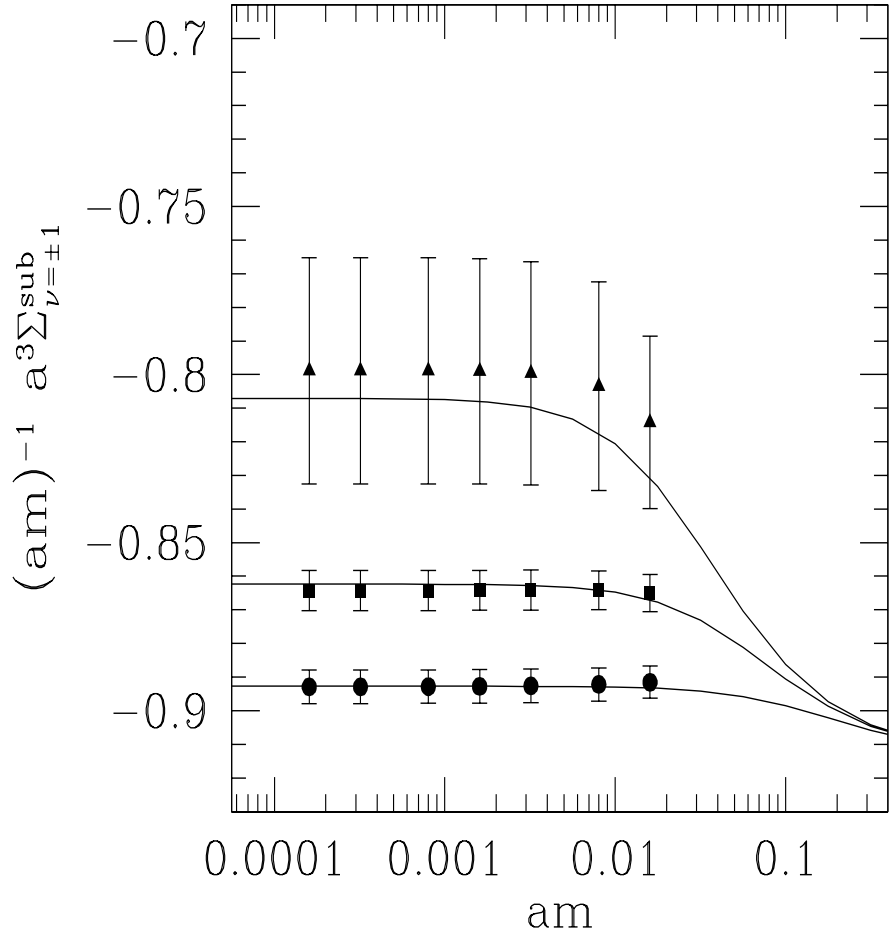, %
width=13cm,height=16cm}
\end{center}
\caption{ \label{fig:pbp} We show the quark mass dependence of 
  the scalar condensate in three volumes, $8^4$ (circles), $10^4$
  (squares) and $12^4$ (triangles). The solid curves represent a fit
  of the data to eqs.~(\ref{eq:sigsub}) and (\ref{qpt}).  }
\end{figure}
 
We show in fig.~1 our results for $a^3\Sigma^{sub}_{\nu=\pm 1}/am$ on
our lattice volumes as a function of bare quark mass. We have 15, 10
and 7 gauge configurations on our $8^4$, $10^4$ and $12^4$ lattices,
respectively. The solid lines are a fit of the data for all volumes
and masses to eqs.~(\ref{eq:sigsub}) and (\ref{qpt}). This fit has only
two parameters, namely the infinite volume, zero quark mass, scalar
condensate $-\Sigma$ and the coefficient of the quadratic divergence.
We find $a^3\Sigma=0.0032(4)$ and $C_2=-0.914(8)$.

Clearly, the formulae derived in $\chi$PT give a very good description
of the numerical data. The infinite volume condensate that we extract
from this fit in physical units is $-\Sigma(\mu\sim 1.5\gev)=-(221^{+8}_{-9}\mev)^3$, up to a multiplicative 
renormalization constant, which has not been computed yet for
Neuberger's operator. We stress that the quoted error on the
condensate is purely statistical. It does not include, for instance,
the expected systematic errors from finite lattice spacing
effects, nor the possible contributions from higher orders in chiral
perturbation theory of $O(F_{\pi} L)^{-2}$. An additional cautoniary 
remark is that the statistics for the largest volume, $12^4$, is rather small as indicated by the large statistical error. We plan to increase
the statistics in the future and include in the analysis also higher topologies, which are more frequent
at larger volumes.

The condensate we obtain is quite close to that reported in ref. \cite{romans} using Wilson fermions and a different method. However, a 
meaningful comparison can only be made when our systematic errors are quantified and the multiplicative renormalization included. 

Recently the authors of \cite{florida} also studied the quark condensate as
a function of quark mass and volume, using Neuberger's operator.
However, a comparison with the predictions of q$\chi$PT in fixed
topological sectors was not attempted and no definite conclusion on
the existence or value of the infinite volume condensate was reached.

According to Random Matrix Theory (RMT), the value of $\Sigma$ may also be
extracted from the distribution of the lowest non-zero eigenvalue
$\lambda_{\rm min}(D_{\rm N})$, defined by the square root of the lowest non-zero eigenvalue of $D_N^\dagger D_N$ at zero quark mass. We have
only gathered a reasonable statistics for the smaller lattice of size $8^4$. 
In the topological charge $\nu =0, 1$ sectors, the corresponding distributions are given by \cite{lmin}:  
\begin{equation}
P_{\nu=0}(z) = \frac{z}{2} e^{-\frac{1}{4}z^2}, \;\;\;\;\; P_{\nu=\pm 1}(z) = \frac{z}{2} I_2(z) \; e^{-\frac{1}{4}z^2}, 
\label{eigendis}
\end{equation}
where $z \equiv \lambda_{\rm min}(D_{\rm N})\Sigma V$.
Recently, the authors of ref.~\cite{florida2} found very good agreement
with these distributions in very small lattices. 
Inserting our value of $a^{3}\Sigma = 0.0032(4)$ in the distribution for zero topology of eq.~(\ref{eigendis}), we get for the
expectation value of this eigenvalue $\langle \lambda_{\rm min}(D_{\rm
  N}) \rangle = 0.135(15)$ (where the error comes from the statistical error
in the condensate), while from our data on the $8^4$  lattice 
(with 41 topology zero configurations) we obtain $\langle
\lambda_{\rm min}(D_{\rm N})\rangle = 0.170(12)$.   
In topology one sectors, the expected value for the 
$\langle \lambda_{\rm min}(D_{\rm  N})\rangle = 0.237(17)$. From the data,
again obtained on the $8^4$ lattice ,with an accummulated statistics of 29 configurations,
we obtain $\langle \lambda_{\rm min}(D_{\rm  N}) \rangle = 0.218(13)$.
In addition,
we generated a sample of eigenvalues according to the distribution of
eq.~(\ref{eigendis}) for $\nu=\pm 1$  with $a^{3}\Sigma = 0.0032(4)$ and a statistics
identical to that of the corresponding simulation. The resulting mean value of
$\lambda_{\rm min}(D_{\rm N})$ and error are fully compatible with
those given by our simulation, indicating that our data do not suffer
from autocorrelation effects. This provides a nice cross check on the value of
$\Sigma$ obtained from our finite-size scaling analysis.

We finally briefly comment on the difficulty in measuring the condensate
in the topology zero sector. This is due to the logarithmic enhancement, which
can be shown to 
originate from the contribution of the single lowest eigenvalue, $\lambda_{\rm min}(D_N)$ to 
the condensate, if the distribution of this eigenvalue is that given by 
RMT in eq.~(\ref{eigendis}). As is clear from eq.~(\ref{eigendis}), 
the distribution of the lowest eigenvalue does not have a gap and 
it is easy to check that the contribution of this single eigenvalue to the condensate, $\Sigma_{\rm min}$, has a logarithmic IR divergence in the sector of zero topology: 
\begin{eqnarray}
\Sigma_{\rm min} = \frac{1}{V}  \;\int d z \; P_{\nu=0}(z) 
\; \frac{2 m}{m^2+ (z/\Sigma V)^2} =  - m 
\Sigma^2 V \log{m \Sigma V} + O(m \Sigma^2 V), 
\label{sigmamin}
\end{eqnarray}
which reproduces exactly the logarithmic dependence in eq.~(\ref{exp}).
We have generated a sample of eigenvalues according to eq.~(\ref{eigendis})
with $a^{3}\Sigma = 0.0032$. In doing so we find that reconstructing the
logarithmic behaviour of the $\nu=0$ condensate requires a statistics much
larger than the one available to us. This is reflected in the results from
our actual simulation where we find that the $\nu=0$ condensate on the
$8^4$ lattice displays very
large statistical errors. For this reason, we have not included these
data in our determination of the scalar condensate.

{\em Conclusion}

Chiral perturbation theory assumes that chiral symmetry is
spontaneously broken in QCD.  Under this assumption it provides, for
small enough values of the quark mass and large enough volumes, the
mass and volume behaviour of the scalar condensate. This behaviour is
determined to lowest order by only one free parameter, namely
$-\Sigma$, the scalar condensate in infinite volume and for zero quark
mass.

Recent developments in lattice QCD have revealed that, contrary to a
long-standing belief, chiral symmetry can be realized on the lattice.
This theoretical advance is connected to the Ginsparg--Wilson relation.
Neuberger proposed a particular operator that satisfies this relation
and we have used this operator in our numerical work.

Although Neuberger's operator is very difficult to treat numerically,
it can be used in practice. In this work we computed the scalar
condensate on lattices of various sizes and for a number of quark
masses, in the regime of chiral symmetry restoration.  The results of
this numerical computation are shown in fig.~1, where we confront our
numerical data with the finite volume and mass behaviour predicted by quenched
chiral perturbation theory.  Obviously, chiral perturbation theory
describes the numerical data well, providing evidence for the 
spontaneous breaking of chiral symmetry.

Although our results are very encouraging, a number of cautionary
remarks have to be made.  The lattices we used are rather small and it
would be desirable to probe the system further on larger lattices. In
addition, it would be important to repeat the calculation at a larger
value of $\beta$ to estimate the lattice spacing effects. Here we were
only able to determine a value of the scalar condensate up to a
multiplicative renormalization constant, which would clearly be needed
for quoting a physical value. Finally, all our results are obtained in
the quenched approximation but, given the complexity of Neuberger's
operator, it would be very difficult to go beyond this approximation.

During the completion of this work, a paper \cite{dehn} using Neuberger's operator to compute the scalar condensate appeared. The data presented in this paper 
are, however, taken in the
strong coupling regime on only one lattice (with small L/$a$=4) and can hence
not directly be compared to our work.

\vspace{0.8cm}

{\em Acknowldegments}
We thank Martin L\"uscher, Massimo Testa and Peter Weisz for many useful
and stimulating discussions and T. Wettig for useful correspondance about RMT. We acknowledge the computer centre at NIC, 
(J\"ulich) and CIEMAT (Madrid) for providing computer time and technical
support. L.L. acknowledges support from the EEC through the TMR 
network EEC-CT98-00169.

\end{document}